# Magnetism in AV$_3$Sb$_5$ (Cs, Rb, K): Origin and Consequences for the Strongly Correlated Phases

Md. Nur Hasan[1], Ritadip Bharati[2], Johan Hellsvik[3], Anna Delin[4,5], Samir Kumar Pal[1], Anders Bergman[6], Shivalika Sharma[7], Igor Di Marco[6,7,8], Manuel Pereiro[6], Patrik Thunström[6], Peter M. Oppeneer[6], Olle Eriksson[6,*] and Debjani Karmakar[6,9,*]

[1] *Department of Chemical and Biological Sciences,*
*S. N. Bose National Centre for Basic Sciences,*
*Block JD, Sector-III, SaltLake, Kolkata 700 106, India*

[2] *School of Physical Sciences*
*National Institute of Science Education and Research*
*HBNI, Jatni - 752050, Odisha, India*

[3] *PDC Center for High Performance Computing,*
*KTH Royal Institute of Technology,*
*SE-100 44 Stockholm, Sweden*

[4] *Department of Applied Physics*
*KTH Royal Institute of Technology*
*SE-106 91 Stockholm, Sweden*

[5] *Swedish e-Science Research Center (SeRC),*
*KTH Royal Institute of Technology,*
*SE-10044 Stockholm, Sweden*

[6] *Department of Physics and Astronomy,*
*Uppsala University, Box 516, SE-751 20, Uppsala, Sweden*

[7] *Asia Pacific Center for Theoretical Physics,*
*Pohang, 37673, Republic of Korea*

[8] *Department of Physics, Pohang University of Science and Technology,*
*Pohang, 37673, Republic of Korea*

[9] *Technical Physics Division*
*Bhabha Atomic Research Centre*
*Mumbai 400085, India*

*Corresponding Authors:
    Debjani Karmakar
    Email: debjani.karmakar@physics.uu.se

    Olle Eriksson
    Email: olle.eriksson@physics.uu.se

**Abstract**

The V-based kagome systems $AV_3Sb_5$ (A = Cs, Rb and K) are unique by virtue of the intricate interplay of non-trivial electronic structure, topology and intriguing fermiology, rendering them to be a playground of many mutually dependent exotic phases like charge-order and superconductivity. Despite numerous recent studies, the interconnection of magnetism and other complex collective phenomena in these systems has yet not arrived at any conclusion. Using first-principles tools, we demonstrate that their electronic structures, complex fermiologies and phonon dispersions are strongly influenced by the interplay of dynamic electron correlations, non-trivial spin-polarization and spin-orbit coupling. An investigation of the first-principles-derived inter-site magnetic exchanges with the complementary analysis of *q*-dependence of the electronic response functions and the electron-phonon coupling indicate that the system conforms as a frustrated spin-cluster, where the occurrence of the charge-order phase is intimately related to the mechanism of electron-phonon coupling, rather than the Fermi-surface nesting.

The V-based kagome stibnite series $AV_3Sb_5$ (A = Cs, Rb and K), or V135 in short, has acquired a centre-stage in the community of topological materials, paralleling twisted bilayer graphene and high $T_c$ cuprates [1,2]. Here, the interplay of strong correlation and non-trivial topology leads to a plethora of exotic phases, harbouring translational, rotational and time-reversal symmetry breaking. The electronic dispersion of $AV_3Sb_5$ displays strongly correlated features near the Fermi energy ($E_F$), e.g. the flatbands of V-$d$ character, van Hove singularities (vHS) at the Brillouin-zone-boundary [3] and non-trivial topological Dirac-crossing of the bands hosting vHS [4]. Their fermiologies are related to the presence of charge bond order (CBO) instability, where the triplet orbital nature ($l = 1$) of the particle-hole wave function bears a contrast to the conventional charge-density wave (CDW) of particle-hole singlets ($l = 0$) [3]. Prior works associated this CBO with the sublattice interference and 3Q-nesting behaviour [5,6], promoting a nematic chiral charge order and spontaneously breaking both the time-reversal and the $C_6$ rotational symmetry [3,7].

Topological Dirac electrons under strong electron correlation and spontaneous symmetry breaking prepare the playground of a co-existing panoply of unconventional and interrelated charge-ordered (CO) and superconducting (SC) phases for V135 compounds [8-10]. The CO-phase emerges at 80-100 K ($T_{co}$) and the SC-phase pertains a critical temperature ($T_c$) of ~1-3 K [9,10]. Spectroscopic imaging detects a complex landscape of superlattice reconstructions related to the CO [1], where the bulk charge-order comprises of a 2×2×4 superlattice with inverse-star-of-David (ISD) pattern in the kagome plane and three consecutive layers of star-of-David (SD) pattern [11,12].

In the scientific community, there is a lack of consensus regarding whether the CO-phase is competing with the SC-phase, trying to gap-out the same band-crossings by similar electronic instabilities or if the former is a precursor of the latter [9,13-16] and also about the pairing symmetry associated with the SC-phase. The available explorations constitute a wide variety of reports on the types of superconducting gaps [17,18] and the corresponding pairing symmetries [19-26].

The presence of local moments and the impact of magnetism on the correlated aspects are not investigated in detail for AV$_3$Sb$_5$. The V-3$d$ states could have been empty in the V$_3$Sb$_5$ network with valence levels V$^{5+}$ and Sb$^{3-}$, unless the alkali-metal (A) donates a single electron to the kagome-net. This single delocalized electron and its hybridization with Sb-5$p_z$ set the stage for the intriguing correlated electron physics. Experimentally, the temperature-dependence of the magnetization is consistent with the Curie-Weiss behaviour at high temperature with an effective magnetic moment of ~0.22 $\mu_B$ per V [10]. In principle, if the V-ions are capable of sustaining a local moment, a spin-bond order with a finite angular momentum should compete energetically with CBO.

Whereas measurements like muon-spin-depolarization in polycrystalline samples divulge negligible local moments [27], for single-crystals, there is a striking enhancement of the internal field below $T_{co}$, implying the presence of a time-reversal symmetry breaking [28]. The most puzzling aspect of AV$_3$Sb$_5$ is the presence of giant anomalous Hall effect (AHE) with anomalous Hall ratio being an order of magnitude higher than that of Fe [29,30]. Conventionally, magnetic topological materials evince an intrinsic AHE related to the Berry phase [31,32]. For these systems, experimental prediction demanded the origin of the AHE to be extrinsic, [29,30] where the electrons undergo an enhanced skew scattering due to the magnetic fluctuations in the triangular spin clusters, as proposed in the "spin cluster" model by Ishizuka and Nagaosa [33]. The prior studies focused on the non-spin-polarized FS-nesting between the electron pockets to the mode-softening at the same value of the displacements ($q$), indicating a structural instability, which disappears under a 2×2×2 reconstruction for KV$_3$Sb$_5$ [34] with a simultaneous exodus of imaginary modes. To the best of our knowledge, the magnetic attributes of these systems have received much less attention. Employing first-principles electronic structure, phonon and time-dependent density functional theory (TDDFT) calculations, we have explored the impacts of electronic correlations and magnetism on their complex non-trivial physical properties.

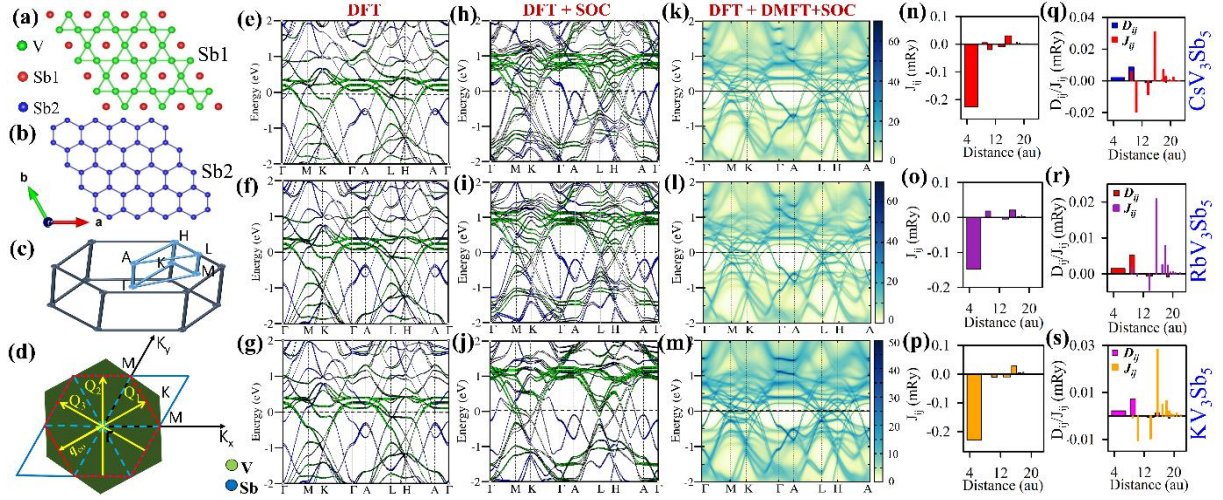

FIG.1. (a) $V_3Sb1$-kagome sublattice and (b) the Sb2 honeycomb sublattice, (c) the hexagonal Brillouin zone (BZ), (d) The 2D BZ (green-filled hexagon), the $k$-space unit cell (blue parallelogram), one ideal FS sheet (red dotted line) and the 3Q-nesting vectors (yellow) with magnitude shown as $q_{co}$, (e-g) The atom-projected band structures for $AV_3Sb_5$ (A = Cs, Rb and K) by spin-polarized DFT, (h-j) corresponding atom-projected band structures with the spin-polarized DFT + SOC. (k-m) corresponding electronic spectral function with the spin-polarized DFT+DMFT+SOC method at 78K, colour scale representing the band-broadening related to quasiparticle lifetimes. (n-p) $J_{ij}$ and (q-s) $D_{ij}$ for $AV_3Sb_5$. In (q-s), the $J_{ij}$ values are in the same plot, except for the first NN.

In the V135 series (symmetry-group *P6/mmm*), the hexagonal A-layer intercalates the quasi-2D network of the $V_3Sb1$-kagome and the Sb2-honeycomb sub-lattices, as in Figs. 1(a)-(b). The corresponding 2D-Brillouin zone (BZ) with an ideal schematic sheet of the Fermi-surface (FS) and nesting vectors (to be discussed below) are presented in Fig 1(c)-(d). With inclusion of spin-polarization, a survey of energetics for the long-range order reveals that while the out-of-plane magnetic coupling is ferromagnetic (FM), as in Table S1 of supplementary materials (SM) [35], the different in-plane spin-configurations, as described in Table S2 of SM [35], are energetically comparable, indicating a frustrated nature for the in-plane magnetism. Fig. 1(e)-

(m) presents a comparison of the spin-polarized atom-projected band dispersions of CsV$_3$Sb$_5$ (CVS), RbV$_3$Sb$_5$ (RVS) and KV$_3$Sb$_5$ (KVS) in the FM configuration, using three different methodologies [35], *viz*., (I) density-functional theory (DFT), (II) DFT including the spin-orbit coupling (SOC) and (III) DFT + dynamical mean-field theory (DMFT) + SOC respectively. Fig. 2 presents the 2D projections of their respective Fermi-surfaces in the [001] plane. The intricate outcomes of this comparative study can be categorized as:

(I) In the collinearly spin-polarized DFT bands (Figs. 1(e)-(g)), the featured signatures constitutes a quasi-2D Γ-centred electron pocket from the Sb-5$p_z$ states, multiple vHS at the M-point from the V-$d_{xy}$ and V-$d_{xz/yz}$ orbitals and the near-$E_F$ Dirac crossing (DC) at the K-point [45]. The corresponding [001]-projected 2D-FS in Fig. 2(a)-(c), are composed of one Γ-centred 2D sheet, one quasi-2D sheet, constituting both the Γ-centred 2D-pocket and 3D-flares near the corners of the BZ, and the 3D M-centred electron pockets related to the vH filling of the V-$d$ levels. These pockets are precisely responsible for the 3Q-nesting as in Figs. 2(a)-(c). In Fig S1 of SM, the band-projected FS sheets are presented [35].

(II) In DFT+SOC, the inclusion of SOC couples the magnetism to the real space, leading to moments along all three spin-axes as in Table S3 [35] of SM. The resultant magnetic moment, being higher than the experimental value, indicates the necessity of a better treatment of the dynamical electronic correlations. The exchange splitting and the associated energy shift push the flat bands away from $E_F$, bring the DCs closer to it, modify the vH filling while also creating new vHS with higher orders [35] and change the area of crossing of the Sb-5$p_z$ bands (blue) near A-point (Figs. 1(h)-(j)). In Figs. 2(d)-(f), the most notable changes occur at the vH-points. The M-centred electron pockets almost disappear for CVS, while a reduction of area is observed for RVS and KVS. The effect of spin polarization is also asserted from the spin-texture-projected FS in Figs. 2(g)-(i).

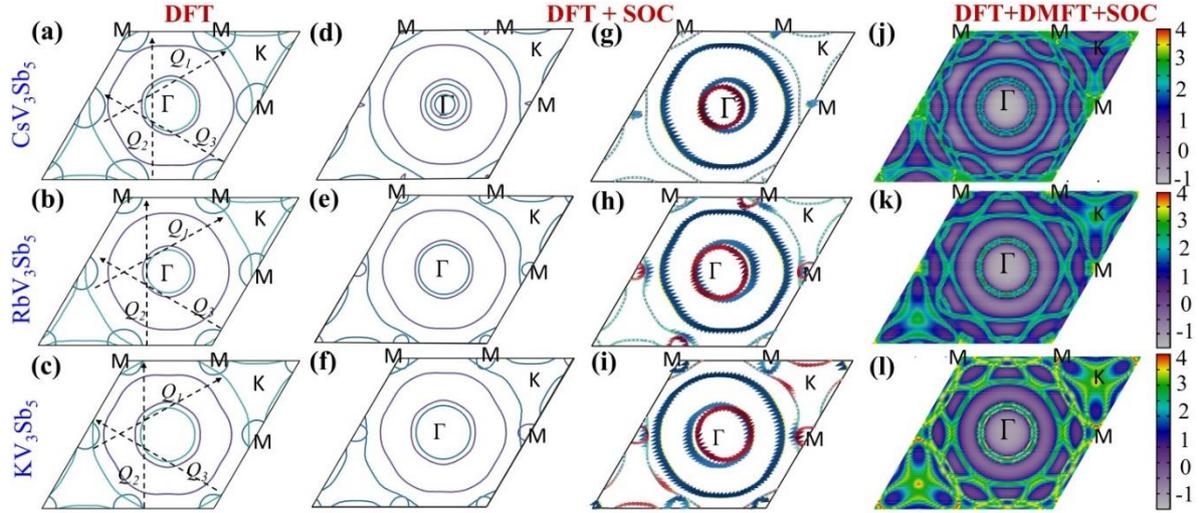

FIG. 2. (a-c) 2D FS of $AV_3Sb_5$ (A = Cs, Rb and K) from spin-polarized DFT. The dotted arrow shows the 3Q nesting behaviour. (d-f) 2D FS and (g-i) spin-textures from DFT + SOC. The red and blue colours signify the spin-up and spin-down projections. (j-l) the 2D FS from DFT+DMFT+SOC. The colour bar represents the Fermi velocity.

(III) For dynamically correlated metals, the state-of-the-art treatment of the local electronic correlations can be obtained by the merger of DFT with the DMFT [46], based on a self-consistent multiband Anderson-impurity model. This model is solved by the fully-relativistic spin-polarised T-matrix fluctuation-exchange (SPTF) solver [47,48], which is efficient for moderately correlated metals with onsite Coulomb correlation U ≤ W/2 (W being the band-width) [49-51]. The electron temperature is kept as 78K ($< T_{co}$). Figs. 1(k)–(m) represents the *k*-resolved correlated electronic spectral functions, where the broadening reflects the quasiparticle lifetimes derived from the imaginary part of the self-energy [51]. Compared to the DFT+SOC, there is an overall flattening of the bands leading to a significant reduction of the band-width with a consequential modification of the low-energy band structure around $E_F$, where the flat V-*d* levels (green bands in Figs 1(h)-(j) below -1 eV) cross the Sb-$5p_z$ electron pockets and the area of the inverted band crossings of Sb-$5p_z$ bands is remarkably reduced. The mass enhancements of the carriers corresponding to the V-*d* levels are in the moderately correlated range, as calculated from the quasi-particle weight extracted from the imaginary part of the self-energy to be ~1.7 (Table 1). In the $v_F$-weighted and smeared FS, as in the Figs. 2(j)–(l), DMFT-induced band renormalizations instigated many additional features, *viz.*, occurrence of additional FS sheets, the negligible areas of the M-centred pockets, highest velocity of the carriers for KVS and the transformation of the K-centred triangular electron pocket into a combination of three pockets situated at around $2\pi/3$ apart. The V-$d$-projected partial spectral functions and the corresponding orbital-projected FS, as presented in Fig. S2 of SM, confirm the contribution of V-$d$ levels towards the dynamical electronic correlation [35].

The significant modifications of the electronic excitations with spin-polarization and strong-correlation suggest the possibilities of their substantial interplay with the elastic properties and thus may have a crucial impact on exotic phases like CO and SC. From the DFT+DMFT+SOC calculations, the local spin and orbital moments of V possess a small but non-negligible value of ~ 0.28 $\mu_B$ and 0.032 $\mu_B$ respectively (Table 1), which is closer to the experimental values of 0.22 $\mu_B$ [9,10].

Understanding of the role of the local moments of V in the itinerant background of the Dirac-like carriers is fundamental to obtain a complete picture of the low-energy magnetic excitations and their impact on both the elastic and inelastic properties of the correlated electron system. Using the Liechtenstein-Katsnelson-Antropov-Gubanov (LKAG) method [52,53], as implemented in fully-relativistic DFT+DMFT+SOC calculations [54,55], the low-lying magnetic excitations are mapped into an effective spin Hamiltonian [54,55] to extract the inter-site exchange parameters like the isotropic Heisenberg exchange ($J_{ij}$) and Dzyaloshinkii-Moriya ($D_{ij}$) interactions. The calculated magnetocrystalline anisotropies suggest that the alignment of the spin and orbital-moments follows an easy-axis pattern, being along the local magnetic $z$-axis, which is also supported by the component-projected spectral functions, as presented in Fig. S3 of SM. Fig. 1(n)-(p) depict the $J_{ij}$ and $D_{ij}$ parameters as a function of the nearest neighbour (NN) distances in the global coordinate system and imply three important facts, in accordance with the experimental scenario. First, the $J_{ij}$ values are an order of magnitude higher for the first NN as compared with the next ones, suggesting the formation of a spin-cluster, as indicated in the AHE experiments [29,30]. Second, the first NN $J_{ij}$s are negative, implying that the V-spins would prefer to align anti-parallel to their NN, which is prevented by the frustration associated with the kagome-lattice-geometry. Third, after the first NN, the $J_{ij}$ and $D_{ij}$ values become comparable in magnitude, as in Figs. 1(q)-(s), suggesting the plausible presence of a spin-spiral like magnetic ground-state. The importance of dynamical correlations in evaluating the exchange properties is reassured from a comparison of the DFT+U and DFT+DMFT results, as seen in Fig S4 of SM [35]. The electronic moments associated to the itinerant Dirac-like and localized flat bands modify the spin-part of the crystal potential, with consequential impacts on the electron-phonon coupling parameter [56,57] and the FS nesting.

The non-magnetic versus the magnetic phonon dispersions of $AV_3Sb_5$ at a temperature of ~ 95 K are plotted in Fig. S5 of SM [35] and Fig. 3 respectively. The non-magnetic phonon dispersions reveal the presence of significant imaginary phonon modes at the M and L-points for both RVS and KVS. CVS, on the other hand, does not display any imaginary mode. The lattice displacements at M and L points of RVS and KVS are displayed in *movie 1* of SM, indicating the substantial role of the interlayer breathing modes. With spin-polarization, CVS continues to be devoid of imaginary modes, as in Fig. 3(a). The absence of the imaginary modes for both magnetic and non-magnetic CVS poses a question towards the interdependence of CO and FS nesting (FSN). For magnetic RVS and KVS, the reduced frequency of the imaginary modes and the diminished areas of the 3D FS pockets (Figs. 2(e)-(f)) iterate the same question. The lattice displacements corresponding to the highest imaginary modes for the spin-polarized RVS and KVS are presented in *movie 2*, where traces of the interlayer breathing modes remain evident. In CVS, the relaxed magnetic structure has the smallest *c*-axis in the V135 series, where the intercalated Cs-ions forms the heaviest and largest-radii alkali-metal layer, restricting the interlayer breathing modes. Pertinent discussions regarding the microscopic analysis are included in SM [35]. Prior studies of the non-magnetic structures related the softened modes at the nesting-$q$-values to the structural instability and CO [34]. Thus, analysis of the CO, FSN and their interrelation with the electron-phonon coupling (EPC) is imperative to acquire a better understanding.

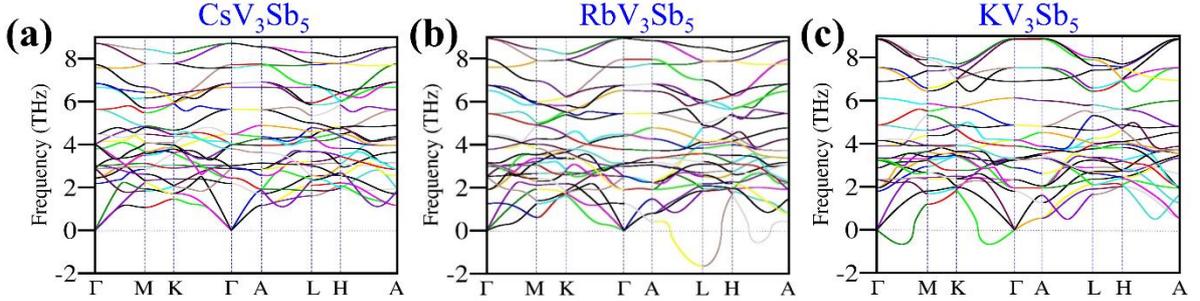

FIG. 3. Phonon dispersion for spin-polarised calculations of: (a) $CsV_3Sb_5$ (b) $RbV_3Sb_5$ and (c) $KV_3Sb_5$. The influence of the magnetic ordering is apparent from the comparison of mode and spin-averaged Eliashberg spectral functions ($\alpha^2 F(\omega)$) in Fig. S7 in SM [35] for magnetic and non-magnetic cases. This in turn will impact the electron-phonon (e-ph) matrix elements, as per the Hopfield relation $\frac{\lambda}{N_\sigma} = D^2/M\omega^2$ with $\lambda$, $N_\sigma$, $D$ and $M\omega^2$ being the e-ph coupling constant, the electronic DOS at $E_F$, the deformation potential and the force constant respectively. By using the density-functional perturbation theory (DFPT), we have calculated the $\lambda$-parameters for both configurations and compared the $\lambda$, $N_\sigma$, $\frac{\lambda}{N_\sigma}$ and the superconducting $T_c$-values from the Allen and Dynes formulation [58] in Table S4 of SM [35]. The $T_c$ values are closer to the experimental values than those of the spin-degenerate ones, as seen in Table 1.

The presence of both FSN and EPC spurs the question of the impact of the respective elastic and inelastic electron-ion scattering on CO. For a 3D system, CO is fundamentally rooted in the electron-ion interaction and thus has an implicit dependence on the electronic response function ($\chi$). Whereas the real part of the response function (Re $\chi$) is connected with the instability causing CO, the imaginary part (Im $\chi$) is coupled to FSN [59]. For systems like CeTe$_3$, where CO is linked to FSN, both Re $\chi$ and Im $\chi$ display a peak at the magnitude of the nesting vector ($q_{co}$) [59]. However, for systems like NbSe$_2$, CO is entirely connected to EPC and the Re $\chi$ and Im $\chi$ are devoid of any peak at $q_{co}$ [59-61]. To investigate the underlying cause of CO in AV$_3$Sb$_5$, using TDDFT [35], we have plotted the *q*-dependence of the interacting ($\chi$) and non-interacting ($\chi_0$) response functions as ω → 0 along Γ-M for both in-plane and out-of-plane magnetizations in Fig. 4(a)-(c). Here, the nesting magnitude ($q_{co}$) is calculated from the approximate schematic in Fig. 1(d). The presence of a shallow non-divergent peak for Re$\chi$ indicates the 3D nature of AV$_3$Sb$_5$ and also predicts the presence of a CO. However, the interdependence of this CO and FSN appears extremely fragile as both $Re\chi(\chi_0)$ and $Im\chi(\chi_0)$ refrains to display a peak at $q_{co}$ for all three systems. Therefore, the role of EPC in CO needs to be examined.

For systems where the instability connected to the CO-phase is linked with the low-frequency acoustic phonon modes, the associated mode-softening almost invariably exhibits an explicit presence of imaginary modes. However, for instabilities linked to the higher-frequency acoustic or optical modes having non-zero frequencies at the zone-centre, it is energetically expensive to obtain a softening after crossing the zero-frequency line. This behaviour is also evident for Bi-2212 cuprates with EPC connected to the $A^{1g}$ optical modes [59-61], as discussed in SM [35].

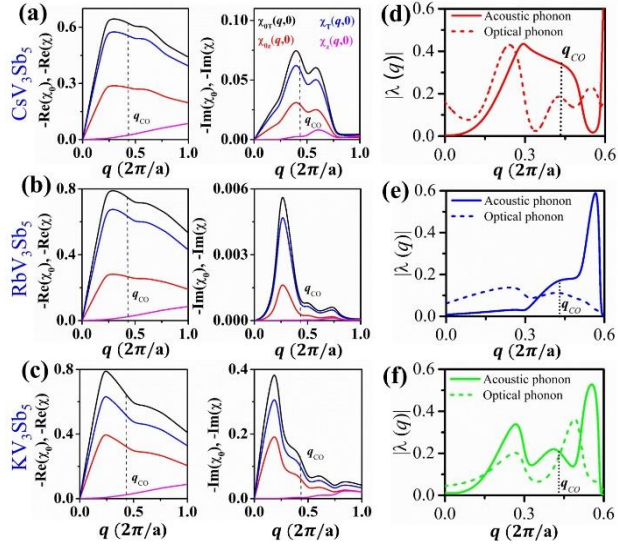

FIG. 4. (a-c) The real and imaginary part of the interacting ($\chi$) and non-interacting ($\chi_0$) response functions for the in-plane ($\chi_T$ and $\chi_{0T}$) and out-of-plane ($\chi_z$ and $\chi_{0z}$) magnetization as $\omega \to 0$, the corresponding colours are displayed as legends; (d-f) The EPC matrix elements $|\lambda(q)|$ for the acoustic and optical (multiplied by four) modes as a function of $q$.

For $AV_3Sb_5$, in spite of the experimental evidences of CO, the absence of imaginary modes in CVS instigated the search for the responsible electron-ion phenomena. Therefore, we calculated the $q$-dependent EPC matrix elements for the acoustic and optical modes corresponding to the largest $\lambda$-parameters (Figs. 4(d)-(f)) along $\Gamma$-M. For acoustic modes of CVS, albeit the peak is away from $q_{co}$, the EPC around $q_{co}$ consists of a broad shoulder region. The optical mode, however, is associated with an explicit peak. This nature of EPC implies a larger contribution of the optical modes towards the structural instability. The shoulder at $q_{co}$ can be attributed to the high-frequency acoustic modes with a non-zero $\omega$, as $q \to 0$. On the contrary, for RVS and KVS, the EPC due to the acoustic modes are more responsible for CO, hosting a corresponding peak near $q_{co}$. Here, softening of the low-frequency acoustic modes leads to the imaginary phonon modes. For RVS, the optical modes may also have a smaller contribution, as indicated by the shallow peak for optical modes at $q_{co}$. The incommensurate FSN in the kagome systems seems to have lesser impact on the CO. The instability of the underlying electronic subsystem is intricately connected to the vibrational instability of the triangular ionic subsystem.

In summary, using first-principles investigations, we have investigated the microscopic origin of the local magnetic moment in the V-based kagome series and its consequences on the electronic and vibrational properties. The three factors, *viz*. the spin and orbital polarization with SOC (SP + OP + SOC), dynamical electron-electron correlations (DEEC) and EPC are demonstrated to have overreaching consequences on the exotic phases, as summarized in the table 2. Rather than the FSN, the phenomenon of CO in this series is observed to have an intricate dependence on the EPC of the system. In conclusion, the incorporation of magnetism and its interplay with the orbital and vibrational degrees of freedom and their impacts on the correlated aspects are capable of explaining several experimental features and simultaneously widen the panorama of exotic physical properties of kagome superconductors.

**Table 1:** (a) Superconducting $T_c$ in K calculated using the EPC in the spin-polarized state from the Allen-Dynes equation, spin moment and orbital moment of vanadium (V), and the average effective mass ($m^*/m_{DFT}$) from DFT+DMFT+SOC.

| System | Superconducting $T_c$ (K) | Spin moment ($\mu_B$) | Orbital moment ($\mu_B$) | Average effective mass ($m^*/m_{DFT}$) |
|---|---|---|---|---|
| CsV$_3$Sb$_5$ | 2.44 | 0.28 | 0.032 | 1.71 |
| RbV$_3$Sb$_5$ | 1.14 | 0.27 | 0.038 | 1.68 |
| KV$_3$Sb$_5$ | 0.95 | 0.28 | 0.039 | 1.72 |

**Table 2**: Summary of the interplay of exotic phases and various possible coupling factors.

| **Exotic Phases** | SP + OP + SOC | DEEC | EPC |
|---|---|---|---|
| **CO** | (a) SP and OP of V-$d$ + SOC + interplay with itinerant electrons produced flatbands, vHS and Dirac-topology<br>(b) reconstructed nesting, vanishing M-point pockets<br>(c) prepared base for incorporation of DECC<br>(d) reduced imaginary modes compared to spin-degenerate theory | (a) Proper treatment of DEEC at 78 K ($< T_{CO}$) produced significant band renormalization<br>(b) reproduced frustrated magnetic environment<br>(c) extracted $J_{ij}$ and $D_{ij}$ explains experimental scenario | (a) Calculated q-dependent EPC has intimate link with CO<br>(b) indicated the vibrational modes leading to instabilities |
| **SC** | (a) $\alpha^2 F(\omega)$ displayed impacts of local moments<br>(b) calculated EPC reproduced experimental $T_c$, by Allen-Dyne formalism. | (a) higher values of effective masses indicated strong correlation<br>(b) extracted $J_{ij}$ and $D_{ij}$ values are consistent with the measured spontaneous symmetry breaking | (a) EPC has relevance for the SC phase and theory reproduces reasonable $T_c$<br>(b) Inelastic electron-ion (EPC) scattering have more impact |

**Acknowledgements**

Financial support from Vetenskapsrådet (grant numbers VR 2016-05980 and VR 2019-05304), and the Knut and Alice Wallenberg foundation (grant numbers 2018.0060, 2021.0246, 2022.0108 and 2022.0079) is acknowledged. The computations were enabled by resources provided by the National Academic Infrastructure for Supercomputing in Sweden (NAISS) and the Swedish National Infrastructure for Computing (SNIC) at NSC and PDC, partially funded by the Swedish Research Council through grant agreements no. 2022-06725 and no. 2018-05973. OE also acknowledges support from STandUPP, the ERC (FASTCORR project) and eSSENCE. MNH acknowledges CSIR (India) for fellowship. IDM acknowledges support from the JRG Program at APCTP through the Science and Technology Promotion Fund and Lottery Fund of the Korean Government, as well as from the Korean Local Governments-Gyeongsangbuk-do Province and Pohang City. IDM and SS also acknowledge financial support from the National Research Foundation of Korea (NRF), funded by the Ministry of Science and ICT (MSIT), through the Mid-Career Grant No. 2020R1A2C101217411. DK acknowledges discussion with Debmalya Chakraborty.

Supplemental Material for

# Magnetism in AV$_3$Sb$_5$ (Cs, Rb and K): Origin and Consequences for the Strongly Correlated Phases

Md. Nur Hasan[1], Ritadip Bharati[2], Johan Hellsvik[3], Anna Delin[4,5], Samir Kumar Pal[1], Anders Bergman[6], Shivalika Sharma[7], Igor Di Marco[6,7,8], Manuel Pereiro[6], Patrik Thunström[6], Peter M. Oppeneer[6], Olle Eriksson[6*] and Debjani Karmakar[6,9,*]

*[1] Department of Chemical and Biological Sciences,*
*S. N. Bose National Centre for Basic Sciences,*
*Block JD, Sector-III, SaltLake, Kolkata 700 106, India*

*[2] School of Physical Sciences*
*National Institute of Science Education and Research*
*HBNI, Jatni - 752050, Odisha, India*

*[3] PDC Center for High Performance Computing,*
*KTH Royal Institute of Technology,*
*SE-100 44 Stockholm, Sweden*

*[4] Department of Applied Physics*
*KTH Royal Institute of Technology*
*SE-106 91 Stockholm, Sweden*

*[5] Swedish e-Science Research Center (SeRC),*
*KTH Royal Institute of Technology,*
*SE-10044 Stockholm, Sweden*

*[6] Department of Physics and Astronomy,*
*Uppsala University, Box 516, SE-751 20, Uppsala, Sweden*

*[7] Asia Pacific Center for Theoretical Physics,*
*Pohang, 37673, Republic of Korea*

*[8] Department of Physics, Pohang University of Science and Technology,*
*Pohang, 37673, Republic of Korea*

*[9] Technical Physics Division*
*Bhabha Atomic Research Centre*
*Mumbai 400085, India*

*Corresponding Authors:
    Debjani Karmakar
    Email: debjani.karmakar@physics.uu.se

    Olle Eriksson
    Email: olle.eriksson@physics.uu.se

**I. Density functional theory (DFT) and DFT+spin-orbit coupling (SOC) Calculations**

The non-spin-polarized and spin polarized plane-wave calculations were performed with the help of projector augmented wave (PAW) pseudopotentials both with and without spin-orbit coupling (SOC). We have used generalized gradient approximation (GGA) with Perdew-Burke-Ernzerhof (PBE) [1] exchange-correlation functional, as implemented in the Vienna ab initio simulation program (VASP) [2]. To examine the role of the Van-der-Waals (VdW) interactions, the semi-empirical dispersion potential representing the dipolar interactions is included within the DFT energy functional by using the DFT-D3 method of Grimme [3]. A Monkhorst-Pack *k*-grid of dimension 5×5×3 is used for sampling the Brillouin zone and the cut-off for the plane wave expansion is kept as 500 eV. The relaxations optimizing the ionic position, cell shape and cell volume have been used to estimate the effects of the structural relaxations on the electronic structure. The energy convergence for all self-consistent field calculations is kept as $10^{-5}$ eV and the structural optimizations were performed using the conjugate gradient algorithm, until the Hellmann-Feynman force on each ion was less than 0.01 eV/Å.

**II. DFT + Dynamical mean-field theory (DMFT) + SOC calculations**

We have used DFT+DMFT+SOC approach to treat the moderately correlated Kagome systems. For strongly correlated systems, the basic task is to find out the actual set of "correlated orbitals" $\{|R, \xi>\}$ with R specifying the Bravais lattice site corresponding to the correlated atom and $\xi$ is the orbital index within the unit cell. In presence of correlation, the total Hamiltonian of the system can be written, after incorporating the Hubbard interaction, explicitly describing the local Coulomb repulsion U, as a Hubbard-like Hamiltonian:

$$H = H_{LDA} + \frac{1}{2}\sum_R \sum_{\xi_1\ldots\xi_4} U_{\xi_1\ldots\xi_4} c^\dagger_{R,\xi_1} c^\dagger_{R,\xi_2} c_{R,\xi_3} c_{R,\xi_4} . \tag{1}$$

Here, $H_{LDA}$ is the Hamiltonian calculated under local density approximation (LDA). With the atomic-like correlated orbitals, the Coulomb parameter $U_{\xi_1\ldots\xi_4}$ can be expressed in terms of the Slater integrals $F^n$ as:

$$U_{\xi_1\ldots\xi_4} = \sum_{n=0}^{2l} a_n(\xi_1\ldots\xi_4) F_n \qquad (2)$$

with

$$a_n(\xi_1\ldots\ldots\xi_4) = \frac{4\pi}{2n+1} \sum_{q=-n}^{n} \langle \xi_1 | Y_{nq} | \xi_3 \rangle \langle \xi_2 | Y_{nq}^* | \xi_4 \rangle \qquad (3)$$

where $\langle \xi_1 | Y_{nq} | \xi_3 \rangle$ and $\langle \xi_2 | Y_{nq}^* | \xi_4 \rangle$ are the integrals over the products of three spherical harmonics.

The Hamiltonian in Eqn.3 represents an effective Hubbard model, the solution of which requires treating a many-body problem. In spectral DFT, after specifying the main observable quantity, we map the original system on to one with lesser degrees of freedom in such a way that the expectation value of the observable is kept intact. In DFT, we define the main observable as the charge density $\rho(r)$ and the one electron Green's function

$$\hat{G}(z) = \left[ (z - \mu)\mathbb{I} - \hat{h}_{LDA} - \hat{\Sigma}(z) \right]^{-1} \qquad (4)$$

here z is the energy in the complex plane, μ is the chemical potential, $\hat{h}_{LDA}$ is the unperturbed Hamiltonian consisting of the hopping term and $\hat{\Sigma}(z)$ is the self-energy operator, being a many-body representative of the electron interactions. In spectral DFT, the main observable is the local Green's function at the site R:

$$\hat{G}_R(z) = \hat{P}_R \hat{G}(z) \hat{P}_R \qquad (5)$$

where, $\hat{P}_R = \sum_\xi |R,\xi\rangle\langle R,\xi|$ is the projection operator in the correlation subspace of **R**. Like LDA or GGA approximations in DFT, the corresponding approximation in spectral DFT is Dynamical mean field theory (DMFT), where the self-energy is assumed to be purely local [4]. This assumption of a local self-energy allows us to concentrate only on single site R and the effect of the other sites can be replaced with a self-consistent electronic bath $\mathcal{G}_0^{-1}(R,z)$. Thus, the atomic site is considered to be embedded in a fermionic bath and that system is treated as a multiband Anderson impurity model. Without knowing the exact Hamiltonian, one can always represent the effective action as:

$$S = -\iint d\tau d\tau' \sum_{\xi_1\xi_2} c_{\xi_1}^\dagger(\tau')[\mathcal{G}_0^{-1}]_{\xi_1\xi_2}(\tau-\tau')c_{\xi_2}(\tau) +$$
$$\frac{1}{2}\int d\tau \sum_{\xi_1\ldots\xi_4} c_{\xi_1}^\dagger(\tau)c_{\xi_2}^\dagger(\tau)U_{\xi_1\ldots\xi_4}c_{\xi_4}(\tau)c_{\xi_3}(\tau) \quad (6)$$

where $\tau$ is the imaginary time for the finite temperature many-body system and the limit of the integral are from 0 to $=\frac{1}{KT}$. So, the problem is now defined and the impurity Green's function $\hat{G}_{imp}(z)$ arises from the dynamics ascribed through the action $S$. With the help of the inverse Dyson equation, an explicit expression for the self-energy operator is obtained after solving the "impurity" problem:

$$\hat{\Sigma}(R,z) = \mathcal{G}_0^{-1}(R,z) - G_{imp}^{-1}(z) \quad (7)$$

The solution of this multiband Anderson impurity problem is obtained with the help of the spin-polarized T-matrix fluctuation-exchange (SPTF) solver [5,6], which is an efficient and reliable impurity solver for moderately correlated electron systems. Once the effective impurity problem is solved and the updated self-energy is obtained, the number of particles will undergo an effective change so that the chemical potential will be updated and thus a new Green's function is obtained as per equation (S2.4). In this way one now obtains a new electronic bath $\mathcal{G}_0^{-1}(R,z)$ by using the inverse Dyson equation:

$$\mathcal{G}_0^{-1}(R,z) = G_R^{-1}(z) + \hat{\Sigma}(R,z). \quad (8)$$

This full process is iterated until a convergent self-energy is obtained. In the next step, this couples with the convergence of electronic charge via the full self-consistent cycle to obtain a complete solution of DFT+DMFT.

All of these calculations have been performed using the relativistic spin polarized toolkit (RSPt)[7-9], which is based on a full-potential description using linearized muffin-tin orbitals.

**III. Calculation of inter-site exchange interactions**

The Heisenberg and DM interactions are calculated using the Lichtenstein-Katnelson-Antropov-Gubanov (LKAG) relation [10,11], where the concerned electronic system is mapped onto a generalized classical Heisenberg model:

$$H = -\sum_{i \neq j} e_i^\alpha \hat{J}_{ij}^{\alpha\beta} e_j^\beta, \quad \alpha, \beta = x, y, z \tag{9}$$

Here the unit vector $e_i$ denotes the local spin direction and the full exchange tensor constitutes interactions like Heisenberg exchange, DM interaction and symmetric anisotropic exchange $\hat{\Gamma}$ by following the relations (for the z-component):

$$D_{ij}^z = \frac{1}{2}\left(J_{ij}^{xy} - J_{ij}^{yx}\right) \tag{10}$$

$$\Gamma_{ij}^z = \frac{1}{2}\left(J_{ij}^{xy} + J_{ij}^{yx}\right). \tag{11}$$

The x and y components can be calculated from similar expressions. All these parameters are extracted from a fully self-consistent converged DFT+DMFT calculation, according to an implementation in reference [12].

**IV. Phonon dispersion of AV$_3$Sb$_5$ (A = Cs, Rb and K)**

The phonon-dispersion curves are calculated with the finite displacement method combining VASP calculations with the Phonopy code. The VASP calculations were performed using the same GGA-PBE exchange-correlation functional with Van der Waals corrections to the functional. The criteria for the minimization of Hellmann-Feynman force is kept as $10^{-6}$ eV/Å and the criteria for convergence of the electronic energy is kept as $10^{-8}$ eV. The dynamical matrices are calculated over a q-point grid of 5×5×5 and the dimension of the supercell is kept as 2×2×2. We successively provide a finite displacement to its atoms by an amount of 0.00048 Å. The temperatures for all the calculations are kept at 95 K as calculated from smearing parameter, which is below the charge-ordering temperature. By using the symmetry operations, the total numbers of such symmetrically distinguishable supercells with atomic displacements are 5. The force constants are generated for each such supercell by using first-principles VASP calculations and these constants, being a signature of the interatomic potential, are collected after using the Phonopy code to generate the dynamical matrix [13]. Diagonalization of this matrix results in the phonon frequencies and thus produces the phonon dispersion curves along the high-symmetry path of the Brillouin-zone.

## V. Superconducting critical temperature ($T_c$) of AV$_3$Sb$_5$ (A = Cs, Rb and K)

The superconducting critical temperature ($T_c$) calculations were performed based on an analysis of Eliashberg's theory proposed by Allen and Dynes [14] with the help of an approximate formula:

$$T_c = \frac{f_1 f_2 \omega_{log}}{1.20} \exp\left(-\frac{1.04\,(1+\lambda)}{\lambda - \mu^*(1+0.62\lambda)}\right) \qquad (12)$$

where $f_1$ and $f_2$ are correction factors that depend on $\lambda$, $\mu^*$, $\omega_{log}$ and $\bar{\omega}_2$

$$f_1 = \left[1 + \left\{\frac{\lambda}{2.46(1+3.8\mu^*)}\right\}^{3/2}\right]^{1/3} \qquad (13)$$

$$f_2 = \left[1 + \frac{\lambda^2\left\{\frac{\bar{\omega}_2}{\omega_{log}} - 1\right\}}{\lambda^2 + \{1.82(1+6.3\mu^*)(\bar{\omega}_2/\omega_{log})\}^2}\right] \qquad (14)$$

and $$\mu^* = \alpha \frac{N}{(1+N)} \tag{15}$$

The value of $\alpha$ is in the range of 0.28 to 0.31. N is the density of states at the Fermi level.

The first inverse moment of $\alpha^2 F(\omega)$ gives the frequency-dependent electron-phonon coupling constant parameter λ from the relation,

$$\lambda(\omega) = 2 \int_0^\infty d(\omega)\, \alpha^2 F(\omega)/\omega. \tag{16}$$

The values of λ and $\omega_{log}$, as extracted from a density functional perturbation theory (DFPT) calculation by using the Quantum Espresso code[15] were directly used in the Allen and Dynes equation. Here, $\omega_2$ is the second moment of the normalized weight function and the general form of the moment is

$$\omega_n = \frac{2}{\lambda} \int d\omega\, \alpha^2 F(\omega) \omega^{n-1}. \tag{17}$$

We have calculated the above integral for $n = 2$ by using $\alpha^2 F(\omega)$ for each value of λ. After getting the value of $\omega_2$, we calculate $\bar{\omega}_2$ as

$$\bar{\omega}_n = (\omega_n)^{1/n} \tag{18}$$

## VI. Time-dependent DFT (TDDFT) for the calculation of interacting and non-interacting response TDDFT Formulation

In TDDFT, the time-dependent Schrödinger equation is mapped onto an effective one-electron problem, where the time dependence is incorporated in the approximation of the exchange-correlation kernel as a function of the explicit time dependence of the exchange-correlation potential and electron-density as:

$$f_{xc}(r, r', t - t') = \frac{\partial v_{xc}(r,t)}{\partial \rho(r',t')} \quad . \tag{19}$$

Calculation of the response function involves a many-body approach towards the solution of the Bethe-Saltpeter Equation (BSE) using the one-body Green's function. In BSE, the dielectric function is written in terms of the bare-Coulomb potential and the interacting response function as [16]:

$$\epsilon^{-1}_{GG'}(q,\omega) = \delta_{GG'} + v_{GG'}(q)\chi'(q,\omega), \qquad (20)$$

where $v(q)$ is the bare Coulomb potential and $\chi$ is the full response function, which is related to the response function $\chi^0$ of the non-interacting Kohn-Sham system as:

$$\chi'(q,\omega) = \frac{v_{GG'}(q)\chi^0_{GG'}(q,\omega)}{1-[v_{GG'}(q)+f_{xc}(q,\omega)]\chi^0_{GG'}(q,\omega)}. \qquad (21)$$

The time-dependent exchange correlated is calculated from:

$$f^{TDDFT}_{xc}(q,\omega) = \frac{\epsilon^{-1}(q,\omega)}{\chi^0_{GG'}(q,\omega)} \qquad (22)$$

The coupled equations (2), (3) and (4) are solved by initially setting $f^{TDDFT}_{xc} = 0$ and then calculating $\chi'(q,\omega)$ and thus $\epsilon^{-1}_{GG'}(q,\omega)$. This value is then utilized in equation (4) to find out the new $f^{TDDFT}_{xc}$. This procedure is repeated until self-consistency is obtained at ω = 0. The response functions are computed after taking care of both interband and intraband contributions.

**VII. Discussion on DFT+U versus DFT+DMFT results for AV$_3$Sb$_5$ (A = Cs, Rb and K)**

Since these systems are moderately correlated metals where dynamical effects are likely to play role, the use of DFT+U may be misleading in calculating the magnetic properties because of its inaccuracy in the treatment of dynamical electronic correlation. Fig. S5 display a comparison of the DFT+U+SOC and DFT+DMFT+SOC results, starting with the small local moments aligned in an FM manner. Whereas the extracted Heisenberg exchange $H_{ij}$ and the corresponding DM interactions from a DFT+U+SOC indicate a strong FM nature, the extracted exchange parameters from DFT+DMFT+SOC resembles the experimentally obtained frustrated AFM spin-cluster-like behaviour. The adiabatic magnon dispersion spectra computed from the exchange parameters extracted from DFT+U+SOC, as presented in Fig. S5(d)-(f), possess a large band-width with the Monte-Carlo-computed magnetization and specific heat indicating a $T_c$ far above room temperature. On the other hand, the adiabatic magnon dispersion with the parameters extracted from DFT+DMFT+SOC, as presented in Fig S5(m)-(o), display a correct band-width and the computed magnetization and specific heat does not result into any sizable $T_c$ due to its frustrated nature. Therefore, the use of DFT + U for such correlated metallic system may lead to contradictory outcomes. A proper treatment of dynamical electron-electron correlation after using DFT+DMFT is an absolute necessity to understand the strong correlation in these systems.

**VIII: Microscopic analysis to understand the absence of imaginary mode in CVS**

The absence of imaginary modes in CVS is intimately tied to its structural chemistry. The spin-polarized structure of CVS, after optimization, has the smallest *c*-axis lattice parameter ~ 8.4 Å among the three systems. The relaxed *c*-axis parameter for RVS and KVS is 8.6 and 8.8 Å respectively. In addition, Cs is more massive with larger ionic radii (167 pm) than Rb (165 pm) or K (152 pm) in its +1 valence state. Therefore, the interlayer breathing modes, leading to the perpendicular movement of the atoms of the V$_3$Sb1 and Sb2 layers, are restricted in case of CVS due to the lesser amount of interlayer space and also due to the presence of massive and larger Cs-ions. To elaborately analyse the vibrational degrees of freedom in CVS, we have introduced the imaginary modes for the phonon-dispersion in CVS after adopting a brute force method of relaxing the stringent criteria of force and electronic convergence, as adopted to obtain the phonon dispersion in the main manuscript. *Movie 3* shows the phonon-dispersion of CVS with such relaxed criteria, where it displays the imaginary modes, mostly linked with low-energy acoustic modes. However, unlike the other two systems, here, the imaginary modes are linked with an in-plane vibration of Cs atoms and not with the interlayer breathing modes. Therefore, the nature of the vibrational degrees of freedom is different in CVS than the other two systems.

This analysis also implies that in order to obtain an out-of-plane interlayer breathing modes, the vibrational energy needed should be much higher. Thus, the interlayer breathing modes, in CVS, are related to the higher frequency long-range damped phonon dispersions near L point, as can be seen in *movie 4*.

**VIII: Interconnection between CO, FSN and EPC**

The origin of CO is closely related to the Peierls-like structural instability seen in systems with reduced dimensionality (mostly quasi-1D linear chain compounds), where the gain in electronic energy exceeds the cost of structural distortion. For such systems, the CO has a purely electronic origin and thus is driven by the Fermi-surface nesting (FSN). There is a metal-insulator transition associated with this type of CO with a logarithmically divergent real part of the electronic response function. The corresponding vibrational mode related to the structural distortion displays a Kohn-anomaly and the respective phonon-softening leads to the occurrence of imaginary modes. For 3D systems, the logarithmic divergence of the response functions becomes weak. If the CO is originated from FSN, both real and imaginary part of the response function should have a peak at the magnitude of the nesting vector $q_{CO}$. In addition, there should be a softening of phonon modes at the same $q$-value. However, there are systems like $NbSe_2$, where the CO is not linked with FSN. The CO instabilities in such systems are originated from EPC. The real and imaginary parts of the response functions are devoid of any peak at $q_{CO}$ and rather the EPC matrix elements display a corresponding peak at $q_{CO}$. This system reveals a phonon damping for a large region of Brillouin zone, in contrast with the sharp Kohn-anomaly seen in quasi-1D systems. To observe a phonon-softening, according to the Grimvall's proposition, the actual experimental phonon-frequency ω(q), inclusive of EPC, can be related to the bare phonon-frequency $\omega_0(q)$ by the relation:

$$\omega(q)^2 = \omega_0(q)^2 + 2\omega_0(q)|g(q)|^2 \text{Re}[\chi(\omega, q)]$$

Presence of an imaginary mode will thus be guided by the criterion,

$$\omega_0(q) < 2|g(q)|^2 \text{Re}[\chi(\omega, q)].$$

However, larger values of EPC (g(q)) may not always ensure the presence of phonon-softening. As is seen for the Bi-2212 cuprates with large EPC, the $A^{1g}$ optical modes connected to the out-of-plane vibrations is related to the structural instability and thus the presence of CO in this system is not associated with any imaginary modes. Wherever the high-energy optical or acoustic modes are linked with CO, it is energetically expensive to have a softening crossing the zero-frequency line. For systems having structural instabilities linked with high-energy (or high-frequency) acoustic or optical modes, the CO may not be associated with any imaginary phonon modes.

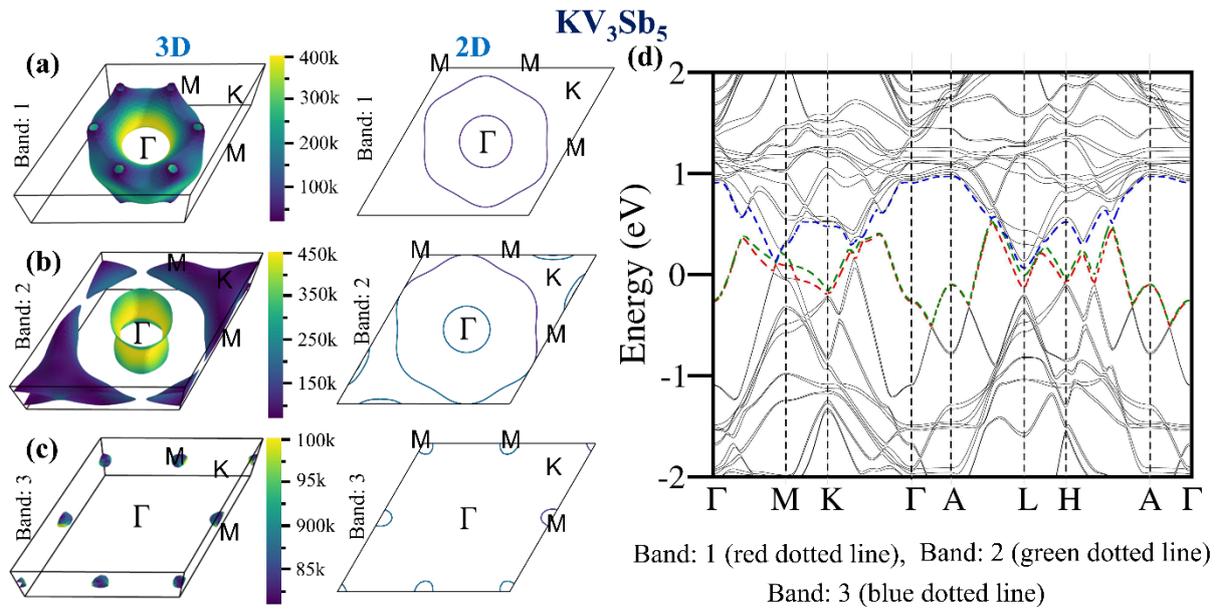

Fig. S1. Band-projected Fermi surfaces of $KV_3Sb_5$ and corresponding band structure. 2D, quasi-2D and 3D FS of $KV_3Sb_5$ corresponding to (a) Band 1, (b) Band 2 and (c) Band 3. (d) Electronic band structure of $KV_3Sb_5$ (Band 1: red dotted line, Band 2: green dotted line and Band 3: blue dotted line). The calculation has been performed computed using the GGA-PBE functional with vdW and PAW potential after incorporating the spin-orbit coupling.

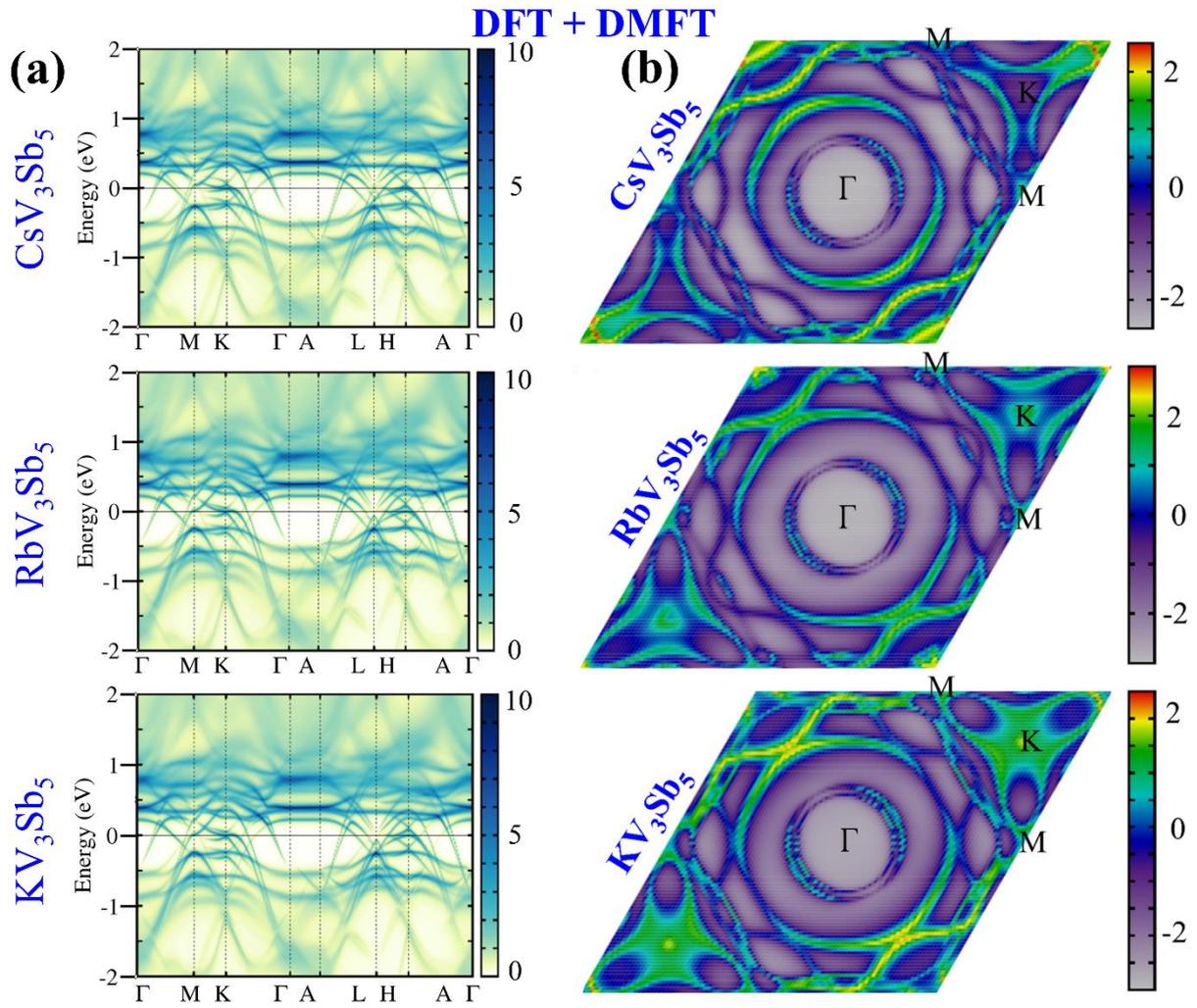

Fig. S2. (a) Partial spectral function and (b) Partial Fermi-surfaces for V-*d* levels in AV$_3$Sb$_5$ (A = Cs, Rb and K) computed with DFT+DMFT+SOC.

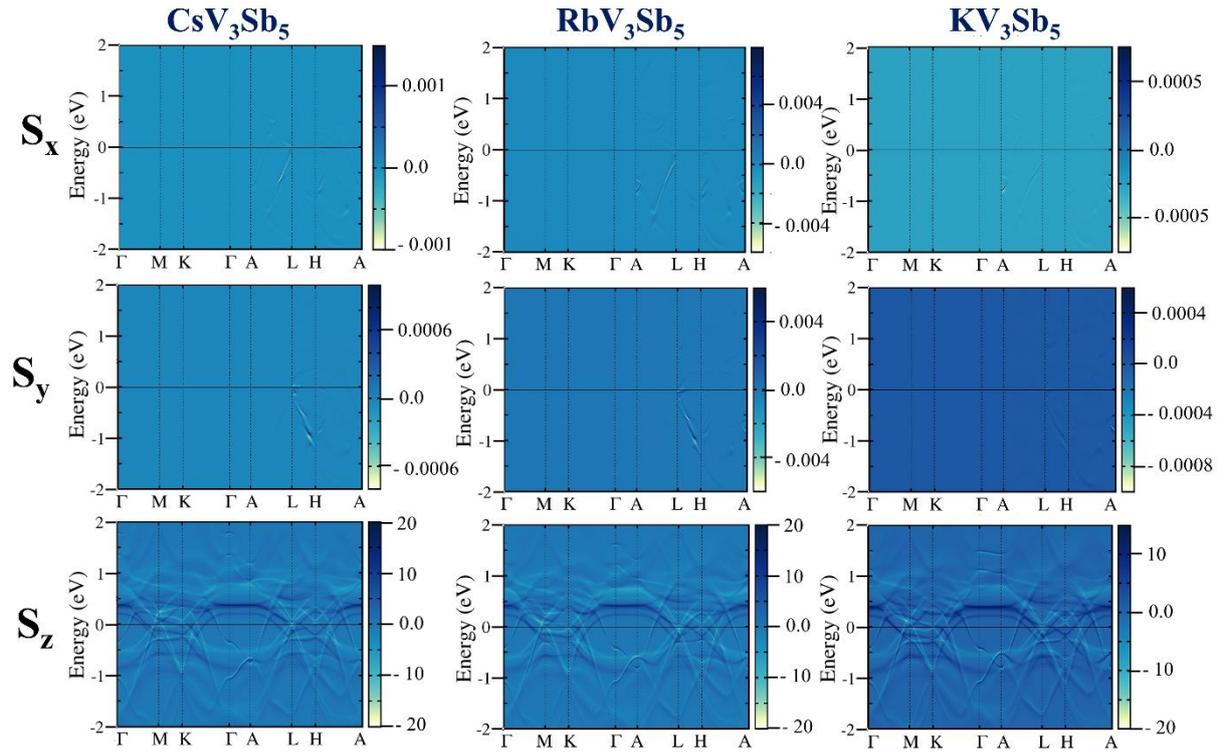

Fig. S3: Spin-projected partial spectral functions of V in $AV_3Sb_5$ (A = Cs, Rb and K). Spin-projected partial spectral functions of V in $AV_3Sb_5$ (A = Cs, Rb and K) computed using the DFT+DMFT+SOC at a temperature of 78 K.

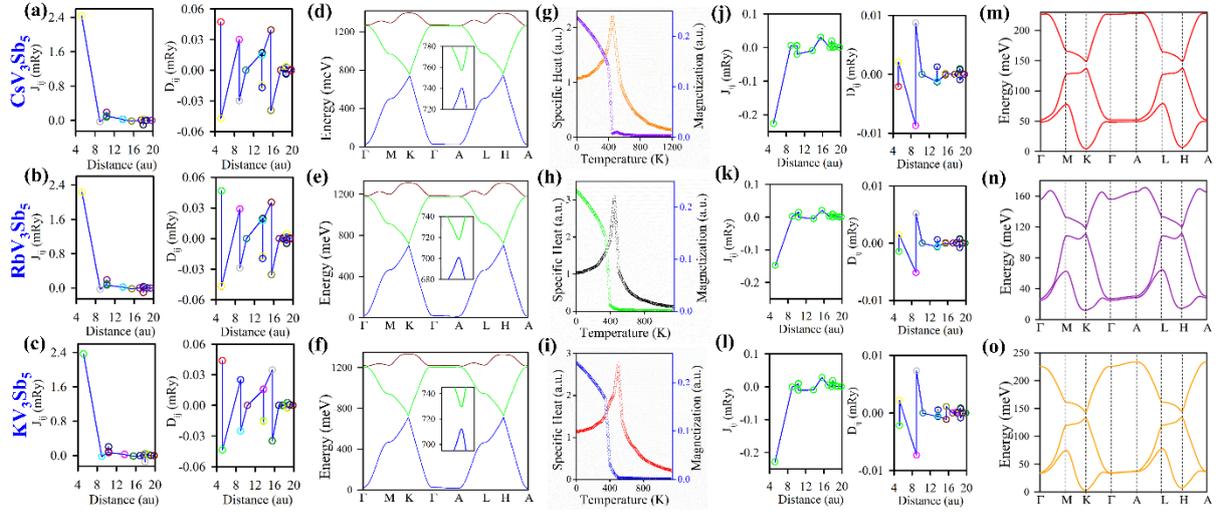

Fig. S4: (a-c) Exchange interactions from DFT+U. The Heisenberg exchange interactions $J_{ij}$ and DM interactions $D_{ij}$ of (a) $CsV_3Sb_5$, (b) $RbV_3Sb_5$, and (c) $KV_3Sb_5$. (d-f) The adiabatic magnon dispersion of (d) $CsV_3Sb_5$ (e) $RbV_3Sb_5$ and (f) $KV_3Sb_5$ weighted with the dynamical structure factor, calculated with the atomistic spin dynamics simulation. (g-i) The specific heat capacity and magnetization curve for (g) $CsV_3Sb_5$ (h) $RbV_3Sb_5$ and (i) $KV_3Sb_5$ computed using the LDA+U. (j-l) Exchange interactions from DFT+DMFT. The Heisenberg exchange interactions $J_{ij}$ and DM interactions $D_{ij}$ of (j) $CsV_3Sb_5$, (k) $RbV_3Sb_5$, and (l) $KV_3Sb_5$. (m-o) the adiabatic magnon dispersions, as obtained from spin-dynamical calculations by using linear spin wave theory of (m) $CsV_3Sb_5$, (n) $RbV_3Sb_5$ and (o) $KV_3Sb_5$.

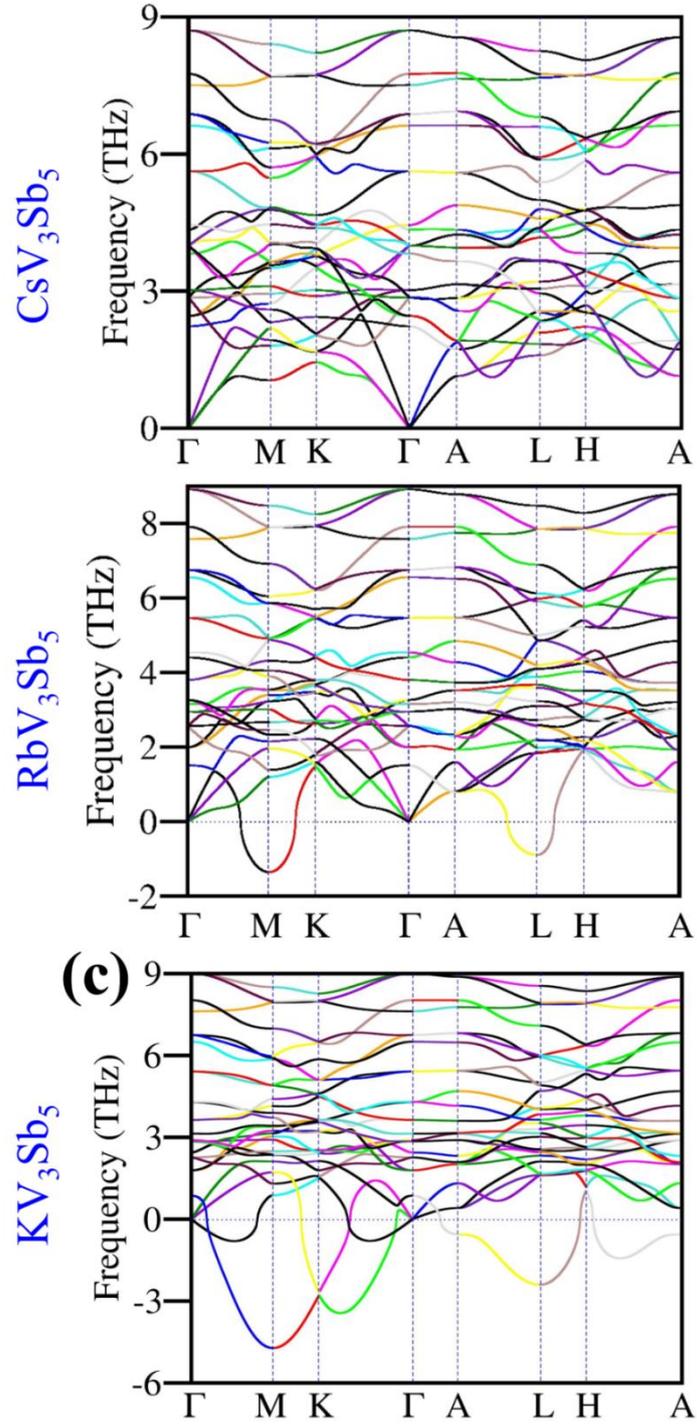

Fig. S5. Non-magnetic phonon dispersion. Non-magnetic phonon dispersion of (a) CsV$_3$Sb$_5$ (b) RbV$_3$Sb$_5$ and (c) KV$_3$Sb$_5$ computed using the GGA-PBE functional with vdW and PAW potential.

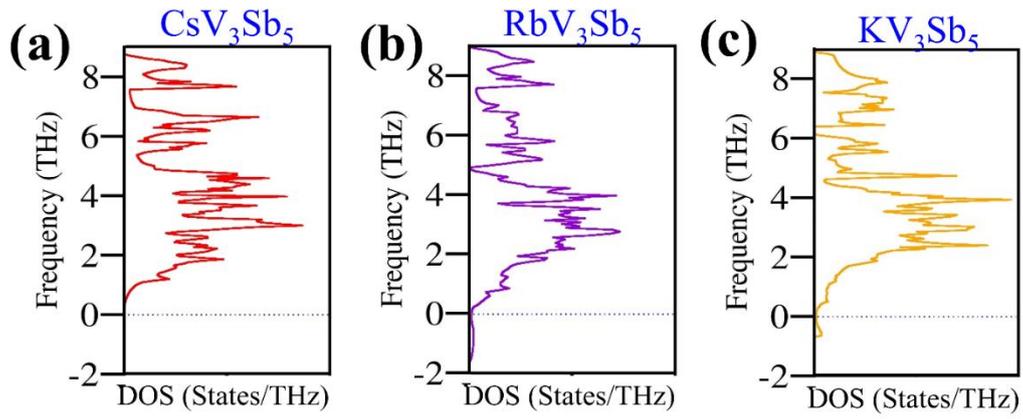

Fig. S6: Magnetic phonon density of states (a) $CsV_3Sb_5$ (b) $RbV_3Sb_5$ and (c) $KV_3Sb_5$ computed using the GGA-PBE functional with vdW and PAW potential.

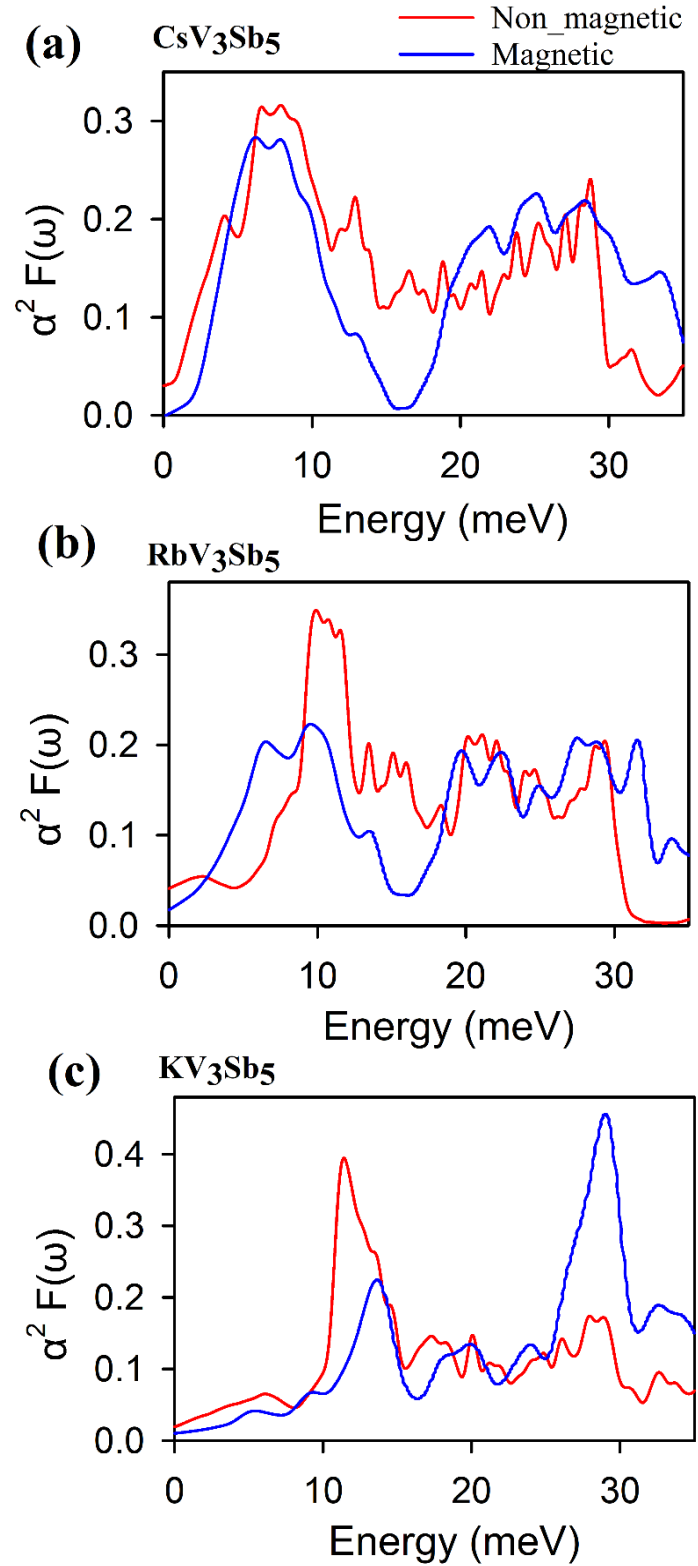

Fig. S7. Eliashberg spectral functions $\alpha^2 F(\omega)$. The Eliashberg functions $\alpha^2 F(\omega)$ of (a) $CsV_3Sb_5$, (b) $RbV_3Sb_5$ and (c) $KV_3Sb_5$ for non-magnetic (red line) and magnetic (blue line) calculations.

Table S1. Energies of FM and out-of-plane AFM configurations by DFT: Comparison of energies between FM and out-of-plane AFM configurations for 1×1×2 supercell of $CsV_3Sb_5$, $RbV_3Sb_5$ and $KV_3Sb_5$ by GGA-PBE with vdW and PAW potentials.

| System | FM (eV) | Out-of-plane AFM (eV) | Energy difference from FM to AFM (meV) |
|---|---|---|---|
| $CsV_3Sb_5$ | -56.553 | -56.172 | 381 |
| $RbV_3Sb_5$ | -54.067 | -53.941 | 126 |
| $KV_3Sb_5$ | -53.825 | -53.764 | 61 |

Table S2. Energies of FM and in-plane AFM configurations by DFT: Comparison of energies between FM and in-plane AFM configurations in the bulk unit cell of $CsV_3Sb_5$, $RbV_3Sb_5$ and $KV_3Sb_5$ by GGA-PBE with vdW and PAW potentials. The configurations (udd) and (udu) imply the spin arrangements (up-down-down) and (up-down-up) respectively.

| System | FM | AFM1 (udd) | AFM2 (udu) |
|---|---|---|---|
| $CsV_3Sb_5$ | -53.353 | -53.301 | -53.305 |
| $RbV_3Sb_5$ | -53.270 | -53.270 | -53.269 |
| $KV_3Sb_5$ | -53.188 | -53.183 | -53.179 |

Table S3. Spin projected moment: Spin projected moment per ion of V of $CsV_3Sb_5$, $RbV_3Sb_5$ and $KV_3Sb_5$ using GGA-PBE functional with vdW and PAW potentials after incorporating SOC.

| System | Moment ($\mu_B$) | | | |
|---|---|---|---|---|
| | $S_x$ | $S_y$ | $S_z$ | $S_{total}$ |
| $CsV_3Sb_5$ | 0.394 | 0.442 | 0.501 | 0.776 |
| $RbV_3Sb_5$ | 0.335 | 0.363 | 0.391 | 0.630 |
| $KV_3Sb_5$ | 0.301 | 0.331 | 0.352 | 0.569 |

Table S4: Non-magnetic and magnetic parameters: Comparison of superconducting critical temperature ($T_c$), electron-phonon coupling constant ($\lambda$) and $N_\sigma(0)$ for non-magnetic and magnetic calculations. $N_\sigma(0)$ is the DOS in states/eV/spin/vanadium.

| System | Non-magnetic | | | | | Magnetic | | | | |
|---|---|---|---|---|---|---|---|---|---|---|
| | $T_c$ (K) | $\lambda$ | $\omega_{log}$ (K) | $N_\sigma(0)$ ($eV^{-1}$) | $\lambda/N_\sigma(0)$ | $T_c$ (K) | $\lambda$ | $\omega_{log}$ (K) | $N_\sigma(0)$ ($eV^{-1}$) | $\lambda/N_\sigma(0)$ |
| $CsV_3Sb_5$ | 1.94 | 1.70 | 33.61 | 6.24 | 0.27 | 2.44 | 1.26 | 82.33 | 6.25 | 0.20 |
| $RbV_3Sb_5$ | 0.47 | 1.06 | 54.78 | 6.22 | 0.17 | 1.14 | 1.40 | 70.76 | 6.24 | 0.22 |
| $KV_3Sb_5$ | 0.88 | 1.14 | 24.98 | 11.23 | 0.27 | 0.95 | 0.63 | 127.55 | 6.23 | 0.10 |